\documentclass[aps,prl,twocolumn,groupedaddress,superscriptaddress,showpacs, showkeys]{revtex4-1}
\usepackage{graphicx}
\usepackage{amsmath}
\usepackage{amssymb}

\newcommand{\ham}[1]{\hat{\mathcal{H}}_{#1}}
\newcommand{\op}[1]{\hat{#1}}
\def\tr{ {\rm{Tr }}}

\begin{document}

\title{Observation of noise-assisted transport in an all-optical cavity-based network}

\author{Silvia Viciani}
\affiliation{CNR-INO, National Institute of Optics, Largo Fermi 6, I-50125 Firenze, Italy}
\affiliation {LENS, European Laboratory for Non-linear Spectroscopy, via Carrara 1, I-50019 Sesto Fiorentino, Italy }

\author{Manuela Lima}
\affiliation{CNR-INO, National Institute of Optics, Largo Fermi 6, I-50125 Firenze, Italy}
\affiliation {LENS, European Laboratory for Non-linear Spectroscopy, via Carrara 1, I-50019 Sesto Fiorentino, Italy }

\author{Marco Bellini}
\affiliation{CNR-INO, National Institute of Optics, Largo Fermi 6, I-50125 Firenze, Italy}
\affiliation {LENS, European Laboratory for Non-linear Spectroscopy, via Carrara 1, I-50019 Sesto Fiorentino, Italy }

\author{Filippo Caruso}
\email{filippo.caruso@lens.unifi.it}
\affiliation {LENS, European Laboratory for Non-linear Spectroscopy, via Carrara 1, I-50019 Sesto Fiorentino, Italy }
\affiliation {Dipartimento di Fisica e Astronomia, Universit\`{a} di Firenze, via Sansone 1, I-50019 Sesto Fiorentino, Italy}
\affiliation{QSTAR, Largo Fermi 2, I-50125 Firenze, Italy}


\begin{abstract}
Recent theoretical and experimental efforts have shown the remarkable and counter-intuitive role of noise in enhancing the transport efficiency of complex systems. Here, we realize simple, scalable, and controllable optical fiber cavity networks that allow us to analyze the performance of transport networks for different conditions of interference, dephasing and disorder. In particular, we experimentally demonstrate that the transport efficiency reaches a maximum when varying the external dephasing noise, i.e. a bell-like shape behavior that had been predicted only theoretically. These optical platforms are very promising simulators of quantum transport phenomena, and could be used, in particular, to design and test optimal topologies of artificial light-harvesting structures for future solar energy technologies.
\end{abstract}


\maketitle

The transmission of energy through interacting systems plays a crucial role in many fields of physics, chemistry, and biology. In particular, the study and a full understanding of the mechanisms driving the energy transport may open interesting perspectives both to improve the process of transferring quantum or classical information across complex networks, and to explain the high efficiency of the excitation transfer through a network of chromophores in photosynthetic systems. Indeed, recently, several experiments on light-harvesting complexes have suggested a possible correlation between the remarkable transport efficiency of these systems and the presence of long-lived quantum effects, observed also at room temperature \cite{qbiobook, Engel2007NAT446, Lee2007SCI316, Collini2010NAT463, Panitchayangkoon10PNAS107, Hildner2013SCI340}. 

Stimulated by these results, a large theoretical effort has been undertaken to study transport mechanisms through a network of chromophores or, more in general, through a complex network, bringing to evidence the active role of noise in energy transport. In fact, it is usually accepted that the uncontrollable interaction of a transmission network with an external noisy environment negatively affects the transport efficiency by reducing the coherence of the system \cite{Zurek2003RMP75}. However, the noise has also been found to play a positive role in assisting the transport of energy \cite{Mohseni2008JCP129, Plenio2008NJP10, Caruso2009JChPh131, Rebentrost2009NJP11, Chin2010NJP12, Caruso2010PRA81} and information \cite{Caruso2010PRL105}, and loss-induced optical transparency has been observed in waveguide systems \cite{Guo2009}. In certain circumstances, the presence of noise can lead to the inhibition of destructive interference and to the opening of additional pathways for excitation transfer, with a consequent increase of the transport efficiency \cite{Caruso2009JChPh131}. This phenomenon has generally gone under the name of noise-assisted transport (NAT). In this context, the role of geometry has been theoretically analyzed in terms of structure optimization \cite{Harel2012JCP136}, in the case of disordered systems \cite{Mohseni2013JCP138}, and also for the proposal of design principles for biomimetic structures \cite{Sarovar2013NJP15}.

Therefore, the possibility to experimentally reproduce NAT effects in purely optical networks with controllable parameters and topology, would allow one to verify the predictions of NAT models in simple test systems. Moreover, the investigation of the system response, when network parameters and noise characteristics are varied, would permit the optimization of specific networks towards maximum transport efficiency \cite{Caruso2014NJP16}. Recently,  an optical platform consisting in a network of coupled cavities has been theoretically proposed as simulator of the basic mechanisms underlying NAT  \cite{Caruso2011PRA83}.

Our goal is to realize a ``proof of principle'' experiment to demonstrate the feasibility of an all-optical cavity-based apparatus and to analyze its potentiality in using well-known classical mechanisms to reproduce the effects of NAT in complex networks. Inspired by  \cite{Caruso2011PRA83}, we have developed a simple and scalable setup where the coherent propagation of excitons in a N-site network is simulated by the propagation of photons in a network of N coupled optical cavities. In particular, our experimental apparatus, entirely based on single-mode fiber-optic components at telecom wavelength (1550 nm) with tunable and detectable parameters, is able to experimentally reproduce the NAT effects in a variety of configurations. 

The choice of single-mode optical fiber components completely removes issues related to matching the transverse spatial mode of the fields and considerably simplifies the alignment of sources, cavities, and detectors, thus allowing us to easily adjust the network size and topology. Working at telecom wavelengths guarantees low optical losses and a low cost of the fiber components, two other important factors for the scalability of the apparatus.
In this context, we chose to use Fiber Bragg Grating (FBG) resonators  \cite{Othons1997RSI68, Hill1997JLT15} as the optical cavities of our network. Such relatively new devices have been mostly employed in sensing applications \cite{Chow2005JLT23, Gagliardi2010SCI330} so far, and are characterized by a straightforward alignment and easy tunability by tiny deformations of the fiber within the Bragg mirrors. These qualities make them ideal for their first application in an optical simulator made of a network of connected tunable cavities.
Each $j$ site of the network can thus be represented by a FBG resonator, and each local site excitation energy corresponds to the resonance frequency of the cavity ($\omega_j$). 

The presence of cavities in our interferometric network, playing the role of chromophores in the light-harvesting complexes, is one of the main differences between the present apparatus and previous optical simulators of quantum walks (QW) as, for instance, \cite{Peruzzo2010SCIENCE329, Broome2010PRL104, Schreiber2011PRL106}. 

We investigate the network response to controlled and measurable amounts of noise. Two kinds of noise are considered here: i) static disorder, that diversifies the energy levels (resonance frequencies) of the different sites (cavities) within the network ($\omega_j \neq \omega_k$, with $j \neq k $), and ii) dynamical disorder (effective dephasing), that introduces random phase perturbations in one or more sites, thus resulting in temporal fluctuations of the corresponding resonance frequencies $\omega_j$'s around their stationary values. In particular, we analyze the dependence of the network transmission as a function of the amount of static and dynamical disorder under different initial conditions, corresponding to maximum or minimum throughput (due to a globally constructive or destructive interference) of the network and also all the intermediate conditions. 

Here we just focus on a few representative results in extreme conditions of a simple geometry to highlight the potential of our platform in reproducing the noise effects in more general networks. In particular, we show that our setup leads to the first experimental observation of a general effect predicted by theoretical models \cite{Rebentrost2009NJP11, Caruso2009JChPh131,  Chin2010NJP12, Caruso2011PRA83}, i.e. a peak in the dependence of the transport efficiency as a function of the amount of dephasing in the network in presence of strong static disorder. 
\begin{figure}[h]
 \includegraphics[width=0.5\textwidth]{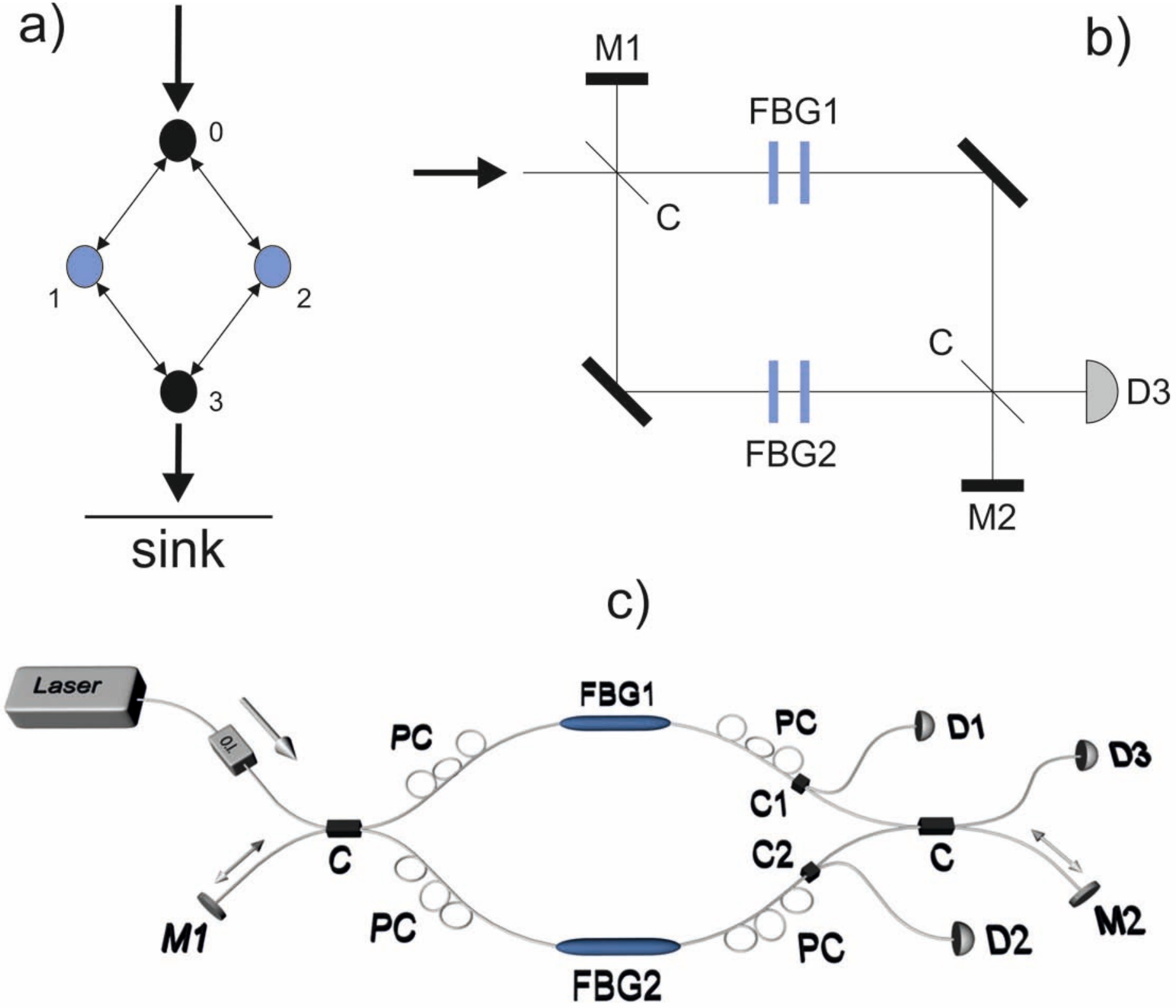}
 \caption{a) Scheme of the 4-site network; b) Schematic setup of our optical platform; c) Detailed experimental setup. Each FBG resonator is inserted in a home-made mounting that allows the fiber cavity to be stressed and relaxed by a piezoelectric transducer, so that its resonance frequency can be tuned. Laser radiation passes through an Optical Isolator (OI) and the input of each resonator is equipped with a fiber polarization controller (PC) to allow only one polarization mode to be confined inside the cavity.}
 \label{fig1}
\end{figure}

\paragraph{Setup.--} To start with a simple topology, we choose to realize a 4-site network, as the one schematically shown in Fig.~\ref{fig1}a. The network comprises an input and an output node and two intermediate ones. It is experimentally reproduced by the optical setup shown in Fig.~\ref{fig1}b, which resembles a Mach-Zehnder interferometer but with two cavities (FBG1 and FBG2) inserted in the two paths. A source (a continuous-wave diode laser, LS, emitting at  1550~nm) injects light of frequency $\omega_\textrm{S}$  into one input port of a first (50:50) fiber coupler. The overall network transmission is measured at one output port of a second (50:50) fiber coupler by means of a photodiode (D3) at the transmission peak of cavity FBG1 (i.e., at $\omega_\textrm{S} = \omega_1$). 
Differently from a simple Mach-Zehnder scheme, the presence of two additional mirrors (M1 and M2) at the normally unused input and output ports of the interferometer partially re-inserts the light reflected and transmitted by the cavities into the network, thus effectively coupling them.
The two FBG resonators correspond to two sites ($j$=1 and $j$=2) of the network, while the role of the other two sites ($j$=0 and $j$=3) is played by the two fiber-optic couplers (C). In this configuration the couplers simulate two sites resonant with the energy of the propagating exciton ($\omega_0 = \omega_3 = \omega_\textrm{S}$), whereas the two cavities have variable local excitation energies ($\hbar\omega_1$ and  $\hbar\omega_2$).  Note that, although optical losses reduce the absolute value of the network transmission, they do not affect its qualitative behavior as a function of noise.

The static disorder of the network can be quantified with the parameter $\Delta x$, which measures the detuning ($\omega_2 - \omega_1$)  between the resonance frequencies of the two cavities in units of their width ($\textrm{FWHM}_{cav} \approx$ 10 MHz).  Two additional fiber couplers (C1 and C2) are therefore used to split a small portion (about 10\%) of the light in each interferometer arm and measure $\Delta x$ from the distance between the transmission peaks of the two cavities, obtained by scanning the laser frequency $\omega_\textrm{S}$.

The output of the system when the two cavities FBG1 and FBG2 are resonant (i.e. when $\Delta x=0$) defines the initial condition for the global interference in the network. For each initial condition of constructive or destructive interference, the system output is then measured for a set of slightly different values of $\omega_2$. In such a way, a statistically significant set of network transmissions (about 14,000) has been obtained for different values of $\Delta x$ and for different global interference conditions. From this dataset, the transmission for a network with static disorder $\Delta x_{0}$ is simply obtained by averaging all the network responses with the same $\Delta x = \Delta x_{0}$. On the other hand, the effect of dynamical disorder (dephasing) is introduced by averaging the network responses corresponding to different values of $\Delta x$. In particular, the transmission of a network with static disorder $\Delta x_{0}$ and dephasing  $\delta x$, is calculated by an average over an interval $\delta x$ around $\Delta x_{0}$.

\paragraph{Model.--}  A numerical model has been developed to describe the dynamics of the proposed 4-site network. Without noise and disorder, the network dynamics can be described by the following Hamiltonian \cite{Caruso2011PRA83}:
\begin{equation}
\ham = \sum_{i} \hbar \omega_{i} \op{a}_{i}^{\dag} \op{a}_{i}  + 
 \sum_{(i,j)} \hbar g_{ij} \left( \op{a}_{i}^{\dag} \op{a}_{j} + \op{a}_{i} \op{a}_{j}^{\dag} \right) \;
\end{equation}
where $\hat{a}_{i}$ and $\hat{a}_{i}^{\dag}$ are the usual bosonic
field operators, annihilating and creating a photon in cavity $i$, $\omega_{i}$ is the cavity resonance frequency, the sum on $(i,j)$ extends over all the connected cavities, and $g_{ij}$ are the coupling rates. Static disorder is then obtained by tuning the $\omega_i$ values, while the additional presence of pure dephasing noise, randomizing the photon phase with rate $\gamma_i$, is described by applying to the quantum state $\hat{\rho}$ the Lindblad super-operator
$\mathcal{L}_{deph}(\hat{\rho}) = \sum_{i} \gamma_{i} \left[ - \left\{ \op{a}_{i}^{\dag} \op{a}_{i}, \op{\rho} \right\} + 2 \op{a}_{i}^{\dag} \op{a}_{i} \op{\rho}
\op{a}_{i}^{\dag} \op{a}_{i} \right]$ with $\{.\,, .\}$ being the anticommutation operation. Finally, the output port (sink), detecting photons and mimicking the photosynthetic reaction center (irreversibly converting the exciton energy into the chemical one), corresponds to $ {\cal L}_{Det}(\hat{\rho}) = \Gamma_{Det} [2 \op{a}_{Det}^{\dag} \op{a}_{k} \hat{\rho} \op{a}_{k}^{\dag}  \op{a}_{Det} - \{\op{a}_{k}^{\dag}  \op{a}_{Det} \op{a}_{Det}^{\dag} \op{a}_{k},\hat{\rho}\} ]$, with $\op{a}_{Det}^{\dag}$ describing the effective photon creation operator in the detector, absorbing photons from the site $k$ (by $\op{a}_{k}$), and $\Gamma_{Det}$ being the coupling rate with the detector. Then, the network transmission is calculated as $\lim_{t \rightarrow \infty} (2\Gamma_{Det} \tr[\hat{\rho(t)} \op{a}_{k}^{\dag} \op{a}_{k}])$ i.e., the rate for photons to reach the sink at sufficiently long times $t$, when it is in a stationary condition, and under the assumption that the network is initially empty (no excitations) and then photons are continuously  fed (by the laser source) into cavity $0$ with a rate $\Gamma_0$. This is modeled (within the Markov approximation) by a thermal bath of harmonic oscillators at a temperature given by the thermal average boson number $n_{th}$ \cite{Caruso2010PRA81}, i.e.
 \begin{eqnarray}
 {\cal L}_{inj}(\hat{\rho}) &=& n_{th}\frac{\Gamma_{0}}{2}[-\{\op{a}_{0} \op{a}_{0}^{\dag},\hat{\rho}\} + 2 \op{a}_{0}^{\dag}\hat{\rho} \op{a}_{0}]
\nonumber \\ &+&(n_{th}+1)\frac{\Gamma_{0}}{2}[-\{\op{a}_{0}^{\dagger}\op{a}_{0},\hat{\rho}\} + 2 \op{a}_{0} \hat{\rho} \op{a}_{0}^{\dag}] \;.
\end{eqnarray}
Then, the dynamical evolution of the state $\hat{\rho}(t)$ is obtained by numerically solving the following differential Lindblad master equation: $\frac{d}{dt}{\hat{\rho}}=-\frac{i}{\hbar} [\hat{H}, \hat{\rho}] + \mathcal{L}_{deph}(\hat{\rho}) +  {\cal L}_{inj}(\hat{\rho}) +  {\cal L}_{Det}(\hat{\rho}) $. Notice that, since our scheme does not involve any non-linear process, a coherent source experiment exhibits the same statistics obtained with a single-photon experiment repeated many times \cite{Amselem2009PRL103}. Moreover, because of the residual asymmetry in the experimental apparatus due to different loss rates in the two cavities, in the model we consider slightly different couplings in the two arms, i.e. $g_{01} \neq g_{02}$ and $g_{13} \neq g_{23}$. This asymmetry can be a further network parameter in a future development of the setup, where the 50:50 fiber-coupler can be easily replaced  by a variable ratio coupler.

\begin{figure}
 \includegraphics[width=0.5\textwidth]{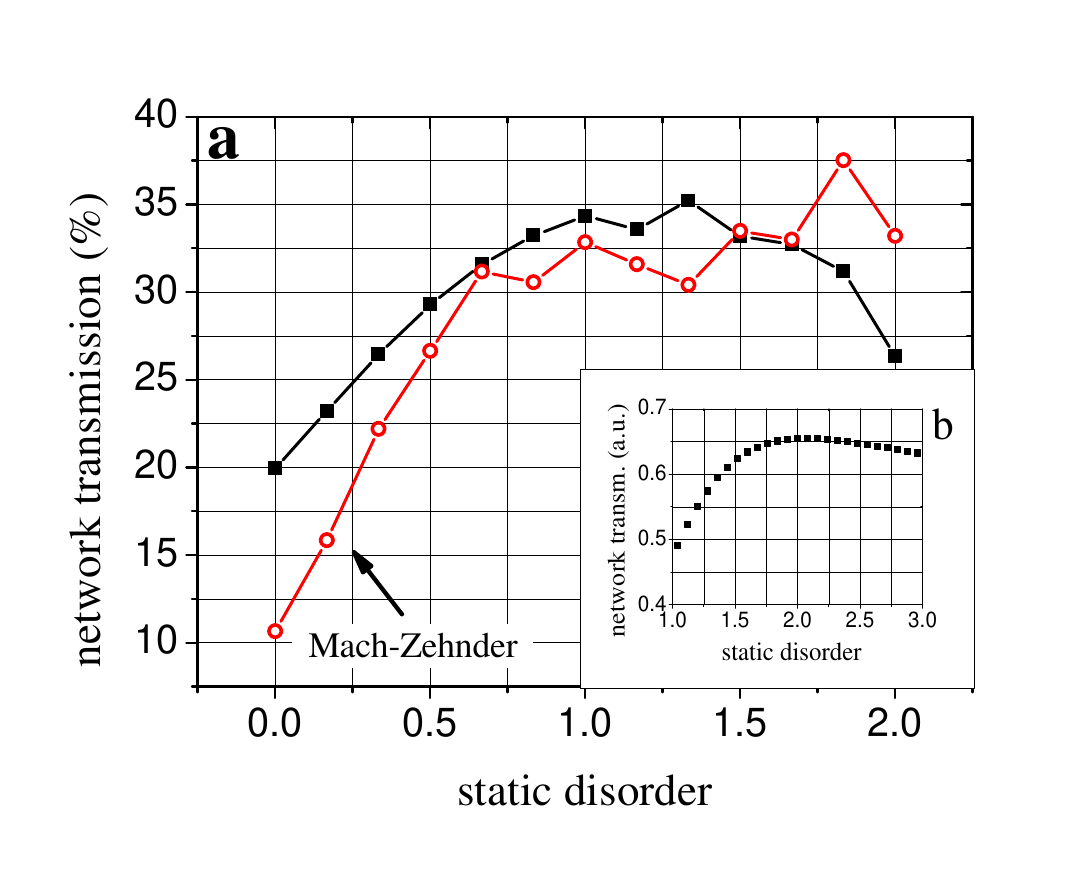}
 \caption{Network transmission $\it{vs}$ static disorder for a system with no dephasing. Comparison between the behavior of the Mach-Zehnder scheme (open circles) and of the complete setup, including mirrors M1 and M2 (filled squares). The initial conditions of the network are chosen such that destructive interference and minimum throughput are enforced in this case. a) Experimental results $\it{vs.}$ $\Delta x$; b) Theoretical results $\it{vs.}$ $(\omega_2- \omega_1)/g_{01}$.}
\label{fig2}
 \end{figure}

\begin{figure}
 \includegraphics[width=0.5\textwidth]{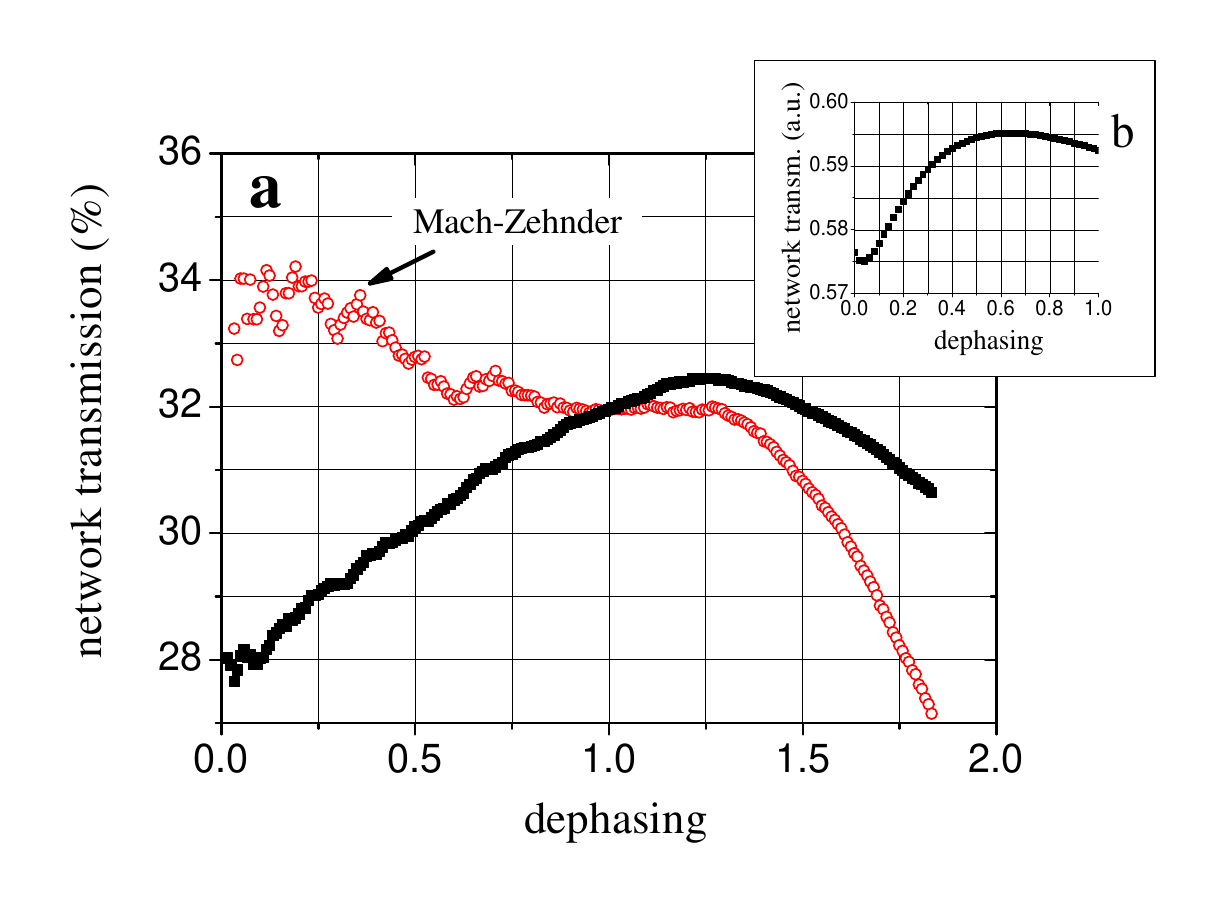}
 \caption{Same as Fig.\ref{fig2}, but network transmission is now plotted as a function of dephasing for a system with static disorder $\Delta x = 2$. a) Experimental results $\it{vs.}$ $\delta x/\Delta x$; b) Theoretical results $\it{vs.}$ $\gamma_2/(\omega_2-\omega_1)$.}
\label{fig3}
 \end{figure}

\paragraph{Results.--} The transmission behavior of the optical setup described above can be intuitively understood in some simple cases. Without mirrors M1 and M2, the scheme reduces to a Mach-Zehnder interferometer with cavities in the two arms. With no disorder, both cavities are resonant with the circulating light (i.e. $\omega_1 = \omega_2 = \omega_\textrm{S}$) and the output of the network should go to zero in the conditions of destructive interference. 
Perfect cancellation of the two symmetric interferometer paths is broken and the network output increases when static disorder is introduced, as the laser source remains resonant with only one cavity but transmission is reduced through the partially resonant one. The network output should then increase to ideally saturate to $1/4$ of the input power when a large static disorder ($\Delta x >1$) makes one cavity completely opaque to the laser radiation. In such conditions of large static disorder, the addition of dephasing noise can recover some transmission through the non-resonant cavity, thus restoring some partial destructive interference between the two interferometer arms, which finally reduces the network transmission again. 
This behavior is clearly observed in the experimental data for the network transmission shown as red open circles in Figs.~\ref{fig2}a and ~\ref{fig3}a, obtained when the mirrors M1 and M2 are removed from the setup. The value of the transmission is normalized to the measured value of the output signal in the conditions of constructive interference and without disorder. The non-zero signal observed without disorder ($\Delta x=0$) is mainly due to asymmetric losses in the two cavities that prevent perfect destructive interference between the interferometer paths. For the same reason, imperfect constructive interference results in a saturation of experimental data at a value larger than the nominal 25$\%$.

The behavior of the optical setup as a function of different kinds of noise becomes much less intuitive when the two mirrors are introduced for simulating the network of Fig.~\ref{fig1}a. The experimental data of Fig.~\ref{fig2}a (black filled squares) indicate an incomplete level of destructive interference even without disorder, due to a partial recycling of the light exiting the interferometer by M2. Increasing the amount of static disorder initially spoils destructive interference, thus increasing the network transmission, until a counter-intuitive decrease of the output (also present in the calculated curve in Fig.~\ref{fig2}b and connected to the complex multi-path interferences) is observed for larger $\Delta x$.  
Differently from the Mach-Zehnder case, introducing dephasing in these conditions of high static disorder does not reduce the transmission. Remarkably, the presence of dynamical disorder initially enhances the network transmission until a maximum is reached for $\delta x \approx 1.25 \Delta x =2.5$ (see Fig.~\ref{fig3}a, black filled squares). 

The described effect is, to the best of our knowledge, the first experimental observation of a peak in the transmission efficiency as a function of dephasing, as predicted by transport models of complex networks \cite{Rebentrost2009NJP11, Caruso2009JChPh131,  Chin2010NJP12, Caruso2011PRA83, Caruso2014NJP16}. According to these theoretical results, the interference between different paths in the network prevents the exciton to use certain channels, thus localizing it on just a few sites and reducing the transport efficiency. By partially destroying the coherence, the dephasing mechanism initially opens additional channels for the transport and allows the exciton to better propagate through the system, resulting in an enhanced transport efficiency. However, when dephasing noise becomes too strong, the transport is expected to be suppressed again. This effect has been considered as an example of the watchdog (quantum Zeno) effect in such noise-assisted transport phenomena \cite{Rebentrost2009NJP11, Caruso2009JChPh131}.
Both these aspects are well reproduced in our optical platform, as clearly illustrated by the qualitative agreement between the experimental curve and the one derived from our theoretical model shown in Fig.~\ref{fig3}b.

If one changes the initial conditions of the network in such a way as to enforce constructive interference and maximum throughput in the absence of disorder, different behavior regimes of the system can be accessed. In this case the setup correctly simulates the reduction of transport efficiency due only to localization in disordered networks because of the difference in the energy levels of different sites \cite{Rebentrost2009NJP11}. With sufficiently high static disorder, a peak in the transmission is again observed when increasing the dephasing.

In conclusion, we have demonstrated the feasibility of an experimental setup of coupled cavities, only based on single-mode fiber optic components, which can efficiently reproduce, in agreement with theoretical models, the performance of transport networks for different conditions of interference, dephasing and disorder.
The use of fiber components gives the possibility to easily increase the numbers of sites and to change the topology of the network. Moreover, also the site coupling can be easily varied by means of fiber couplers with different ratios. Similar systems might become relatively simple and scalable simulators with several controllable parameters that allow a complete investigation of the dependence of the transfer efficiency as a function of both the network characteristics and the noise features \footnote{A complete analysis of all the different configurations that can be efficiently simulated by our apparatus is currently in progress and will be discussed elsewhere}. 

This apparatus represents just the first step towards the realization of optical simulators of quantum transport phenomena. A true computational advantage of such simulators could be achieved when non-classical light states are used in combination with coincidence detection, as in the case of Boson sampling \cite{Scott2011}. 

Unlike any real biological sample, our purely optical setup has the remarkable advantage of allowing a complete control over all the fundamental system parameters, e.g. tuning the coupling rates between different chromophores, playing with energetic static disorder, and adding an optimal amount of noise to enhance transport. Furthermore, our platform, being sufficiently re-configurable, will allow us to create and test different network topologies. These tasks are unfeasible with natural photosynthetic systems, and, also in the case of artificial photosynthetic molecules, the remarkable cost of producing them does not allow one to freely manipulate and explore many different configurations. Therefore these results point towards the possibility of designing optimized structures for transport assisted by noise, that might also be used for future and more efficient solar energy technologies.
After the submission of this manuscript, another group \cite{Kassal2015arxiv} has reported the observation of noise-assisted transport in an optical network, but showing only the increasing profile of the bell-like shape behavior of the transmission efficiency as a function of the dephasing rate.

\begin{acknowledgments}
The authors gratefully acknowledge fruitful discussions with P. Scudo and R. Fusco. This work was supported by the Future in Research (FIRB) Programme of the Italian Ministry of Education, University and Research (MIUR), under the FIRB-MIUR grant agreement No. RBFR10M3SB, and performed in the framework of the ENI contract No. 3500023215. The work of F.C. has been also supported by a Marie Curie Career Integration Grant within the 7th European Community Framework Programme, under the grant agreement QuantumBioTech No. 293449.
\end{acknowledgments}

\bibliography{mioBIB1}

\end{document}